\documentclass[%
 reprint,
 amsmath,amssymb,
 aps,
]{revtex4-1}

\usepackage{graphicx}
\usepackage{dcolumn}
\usepackage{bm}

\begin{document}


\title{Ultrafast transport and relaxation of hot plasmonic electrons in metal-dielectric heterostructures}

\author{Ilya Razdolski}
\altaffiliation[]{Present address: FELIX laboratory, Radboud University, 6525 HP Nijmegen, The Netherlands}
\email{e-mail: i.razdolski@fhi-berlin.mpg.de}
\affiliation{Fritz Haber Institute of the Max Planck Society, 14195 Berlin, Germany}
\author{Alexander L. Chekhov}
\altaffiliation[]{Present address: Department of Physics, Free University Berlin, 14195 Berlin, Germany}
\affiliation{Department of Physics, Moscow State University, 119991 Moscow, Russia}
\author{Alexander I. Stognij}
\affiliation{Scientific-Practical Materials Research Center of the NASB, 220072 Minsk, Belarus}
\author{Andrzej Stupakiewicz}
\affiliation{Faculty of Physics, University of Bialystok, 15-245 Bialystok, Poland}

\date{\today}

\begin{abstract}
We analyze ultrafast electron dynamics in the time domain upon optical excitation of propagating surface plasmon-polaritons (SPs) in metal-dielectric heterostructures. Developing a kinetic model where both local and non-local electron relaxation in metals are included, we identify relevant timescales and extend the existing non-equilibrium electron dynamics framework onto the case of collective electronic excitations. The experimental data obtained in two distinct series of pump-probe measurements (with varied pump wavelength and angle of incidence) demonstrate SP-driven, one order of magnitude enhanced efficiency of the hot electron generation and 
the fourfold (up to 200 fs) slowdown of their non-local relaxation 
at the SP resonance. We discuss the perspectives of the SP-enabled manipulation of the non-equilibrium electron population lying at the crossover of photonics and ultrafast spintronics.

\end{abstract}

\pacs{Valid PACS appear here}
\maketitle



Ultrafast dynamics of electronic excitations is of growing fundamental interest in modern physics. Recently, non-thermal possibility of spin control in magnetic media has been discovered, featuring excitation of the electronic subsystem by ultrashort laser pulses \cite{StupakiewiczNature17}. Strong enhancement and nanoscale localization of the optical spin excitation in resonant metal-dielectric bilayers were demonstrated, owing to the surface plasmon-polaritons \cite{ChekhovNanoLett18}.
It is now widely accepted that femtosecond electron dynamics after an ultrashort optical excitation
cannot be accurately described within the thermal equilibrium models. Instead, non-equilibrium electron distributions should be considered on the timescale of a few hundred of femtoseconds, depending on the particular metal and excitation photon energy \cite{FannPRB92,GroeneveldPRB92,SunPRB94,GroeneveldPRB95,DelfattiPRB00,RethfeldPRB02,MuellerPRB13}. Alternative models consider non-thermal distributions of phonons \cite{WaldeckerPRX16,OnoPRB17}. Simultaneously, small penetration depth of the optical field in metals ensures inhomogeneous excitation profile in thicker films, thus enabling in-depth transport of hot laser-excited electrons. The importance of these non-local processes has been shown for the heat transport \cite{LejmanJOSAB14} and laser-induced magnetization dynamics \cite{MalinowskiNatPhys08,BattiatoPRL10,EschenlohrNatMater13,SchellekensAPL13,WieczorekPRB15,BergeardPRL16,RazdolskiJPCM17}.

Yet, required for the rigorous description of ultrafast electron dynamics, simultaneous consideration of these two effects has been limited to the demonstration of ballistic transport of hot non-thermal electrons across large ($\sim 10^2$~nm) distances \cite{MelnikovPRL11,AlekhinPRL17}. There, laser-induced bunches of hot electrons at elevated energies quickly move away from the excitation volume, requiring
detection by optical methods with femtosecond temporal resolution.
Notably, this electron dynamics can be sufficiently well understood within a single-particle picture, whereas
the role of collective excitations (i.e. plasmons) in the ultrafast non-local electron dynamics remains unexplored. Scarce time-resolved investigations of the electron dynamics at surface plasmon (SP) resonances have been largely limited to the metal nanoparticles \cite{BigotPRL95,LamprechtPRL00,LehmannPRL00,VoisinJPCB01,SaavedraACS16}, thus minimizing the electron transport contributions and facilitating faster thermalization through Landau damping \cite{KawabataJPSJ66,Khurgin15}. In other works, the characteristic timescale of the ballistic transport of laser-excited electrons was found too fast to be analyzed on equal footing with the local dynamics of the electron distribution \cite{RotenbergPRB09}.

In this Letter, we employ the pump-probe technique with femtosecond temporal resolution to analyze both local and non-local electron dynamics in the vicinity of the SP resonance on Au grating. Continuously varying the pump wavelength across the resonance, we register transient pump-induced variations of transmittance and outline relevant processes for the electron dynamics. 
We observe significant increase of the hot electron transport timescale and clearly attribute it to the excitation of the SP resonance.
Showing a high degree of consistency in the determination of the SP and hot electron lifetimes, our results elucidate the previously unexplored role of SP excitation in the non-equilibrium electron dynamics.

\begin{figure}[t]
    \centering
    \includegraphics[width=0.95\columnwidth]{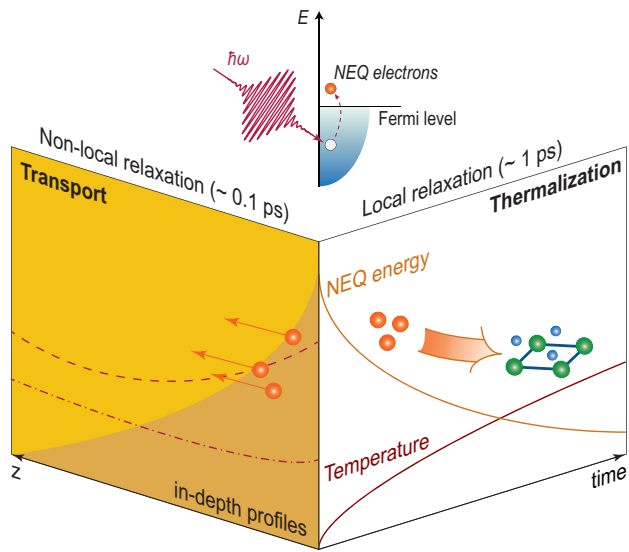}
    \caption{Overview of the electron dynamics upon ultrafast laser excitation of a thick metal film. Left: Non-thermal population (the shaded area) is created within the optical penetration depth $\zeta$. Superdiffusive transport of hot electrons leads to the quick in-depth redistribution (dashed and dash-dotted lines) and non-local relaxation of their population. Right: Thermalization of the hot electrons (local relaxation) results in the temperature rise across the metal. 
    The inset illustrates the energy flow from the non-equilibrium electrons (orange) to the thermalized electrons (blue) and lattice (green).
    The top inset shows the laser-induced creation of the non-thermal electrons population.
    }
    \label{figs:over}
\end{figure}

Femtosecond laser excitation of a metal results in the generation of hot electrons 
above the Fermi level (Fig.~\ref{figs:over}, inset). The spatial distribution of these primary electrons is governed by the optical excitation profile, typically (for plasmonic metals (PM) in the visible or near-infrared spectral range) on the order of $20$ nm. As such, in thicker metallic films the inhomogeneity of the excitation enables subsequent in-depth redistribution of the hot electron population (Fig.~\ref{figs:over}, left), which can be described within the appropriate transport models. In particular, superdiffusive and ballistic transport mechanisms have been argued to play a key role on the ultrafast timescale \cite{BrorsonPRL87, BattiatoPRL10,BattiatoPRB12}. In Au, large mean free path values ($\gtrsim100$~nm) for hot electrons were found \cite{KrolikowskiPRB70,BrorsonPRL87,JuhaszPRB93,MelnikovPRL11}, thus indicating the ballistic character of the electron transport with the Fermi velocity $v_F\sim1$~nm/fs. 

The laser-generated non-equilibrium population of hot electrons decays by virtue of the electron-electron and electron-phonon scattering (Fig.~\ref{figs:over}, right). Experimentally, the importance of non-equilibrium electron distribution for the transient optical response on the ultrafast timescale has been established in various PM systems \cite{GroeneveldPRB95,HohlfeldAPB97,DelfattiPRB00,RotenbergPRB09}. 
The lifetime of the non-equilibrium electron distribution $\tau_{N}$ in the pump-probe experiments was found to be $\sim0.5-1$~ps. We emphasize that this $\tau_{N}$ should not be confused with the average single electron lifetime $\tau_e$ (or inverse electron scattering rate $\gamma^{-1}$) which for the electrons at energies $E\sim 1$~eV is $\sim 10-30$~fs, as predicted by the Fermi liquid theory and measured in the photoemission experiments \cite{SchmuttenmaerPRB94,KnoeselSurfSci96,BauerProgress15}. This striking difference is related to the electron multiplication, resulting in the temporary increase of the number of non-thermal electrons and the lifetime of their population.
Until the electron distribution is relaxed to the Fermi-Dirac one, the two-temperature model description, utilizing well-defined electron and lattice temperatures
cannot be employed. Eventually, 
the electron-phonon system relaxes towards thermal equilibrium at an elevated temperature (Fig.~\ref{figs:over}). All these dynamic processes make their way into the transient optical properties of PM systems and can be identified from the time-resolved optical experiments.

Analysing the typical timescales above, it becomes clear that in $\lesssim 100$~nm-thick PM films, ballistic transport of hot laser-excited electrons occurs on much faster timescale than the electron thermalization. Thus, although both electron transport and thermalization contribute to the transient variations of the optical properties in an intricate way, the disparity of the timescales enables the disentanglement of these effects in the experiments. 
Spitzer {\it et al.}~\cite{SpitzerPRB16} reported $\tau_{\rm S1}\approx 220-300$~fs, $\tau_{\rm S2}\approx 940$~fs in a $130$~nm-thick Au layer, claiming the relaxation of the hot electrons within $30-80$~fs and therefore excluding it from consideration. 
However, previous studies of hot electrons in PMs  \cite{SunPRB94,GroeneveldPRB95,DelfattiPRB00,RotenbergPRB09} suggest much longer relaxation times of the non-thermal electronic population (up to $1$~ps).
This underestimation of the effective lifetime of hot electrons and its lack of sensitivity to the excitation of SP resonance has prompted the authors to look for alternative 
explanations.

\begin{figure}[t]
    \centering
    \includegraphics[width=0.95\columnwidth]{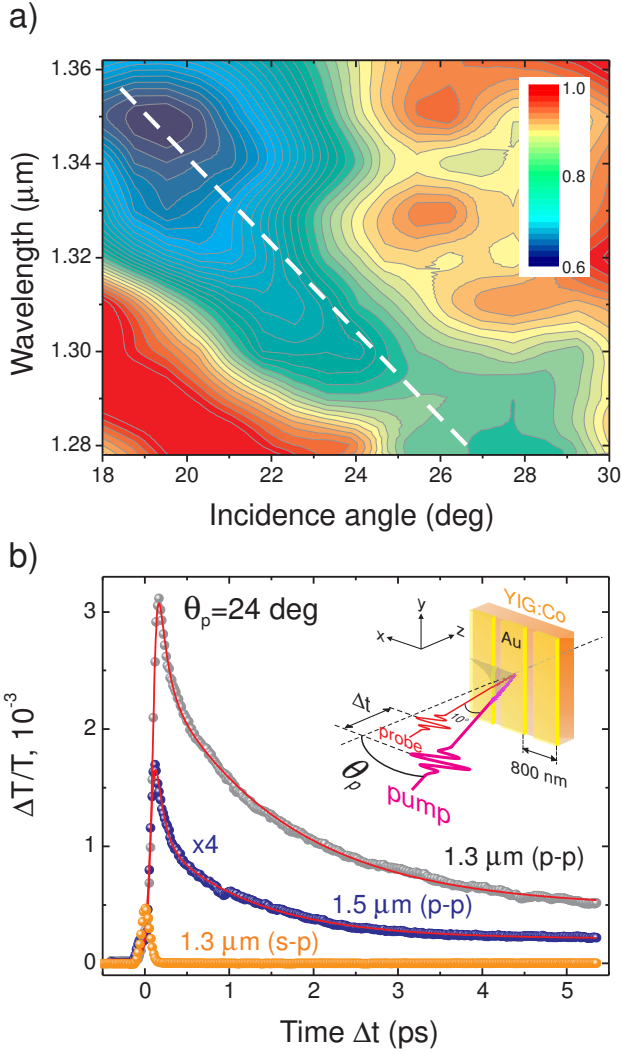}
    \caption{
    a) False color transmittivity map of the plasmonic Au-garnet grating. The low transmittivity area illustrate the SP dispersion, as highlighted with the white dashed line. The spectral width of the incident radiation is on the order of 50~nm.
    b) 
    Time-resolved transmittivity variations for the SP-resonant case (p-polarization, $\lambda_{\rm p}=1.3~\mu$m) and two non-resonant cases.
    The red solid lines are the fit curves using Eq.~(2). The inset: schematics of the pump-probe experiments.
    }
    \label{figs:exp}
\end{figure}

To reveal the role of the SP excitation in the ultrafast hot electron dynamics, in this work we employed 40~fs-long laser pulses for studying transient transmittance of a periodically corrugated Au/Co-doped yittrium iron garnet (YIG:Co) magneto-plasmonic crystal. This ferrimagnetic dielectric garnet is a promising material for magnetic recording applications \cite{StupakiewiczNature17} whereas photonic excitations can be utilized for the nanoscale localization of the laser-excited spin dynamics \cite{SchmisingNJP15,LiuNanoLett15,ChekhovNanoLett18}. The metallic grating enables 
free-space excitation of propagative SPs 
\cite{Belotelov11,Belotelov13,Chin13,PohlPRB12,KrutyanskiyPRB15,RazdolskiACS15,ChekhovPRB16,SpitzerPRB16,ChekhovNanoLett18}, according to the well-known phase matching condition $k_{\rm SP}=k_0\sin\theta+mk_G$.
%
Here $k_{\rm SP}$ is the SP wavevector, $k_{\rm in}=k_0\sin\theta$ is the in-plane component of the free-space photon momentum, $k_G=2\pi/b$ is the quasiwavevector of a grating of period $b$, and $m$ is an integer. 
Experimentally, we studied a magneto-plasmonic Au-YIG:Co crystal with $b=800$~nm, gap width 100~nm, and Au thickness $d=50$~nm
\cite{ChekhovNanoLett18}. Transmittance angular spectra taken at a series of fundamental wavelengths in Figure~\ref{figs:exp},a illustrate the SP dispersion in the near-IR range. 
%
%
%
In the first series of time-resolved measurements, the pump (fluence $\sim1$~mJ/cm$^2$) angle of incidence was fixed at $\theta_p=24$~degrees, and its wavelength $\lambda_p$ was tuned in the near-IR spectral range around the SP resonance at the Au/garnet interface at about $\lambda_{\rm SP}\approx1.3$~$\mu$m.
The choice of this wavelength is motivated by the resonant nature of the laser-induced magnetization switching in YIG:Co \cite{StupakiewiczNature17}.
The typical time-resolved trace of the transmittance variations $\Delta T(t)$ of the delayed, weak probe beam at $800$~nm wavelength is shown in Fig.~\ref{figs:exp},b. 

\begin{figure}[t]
    \centering
    \includegraphics[width=0.95\columnwidth]{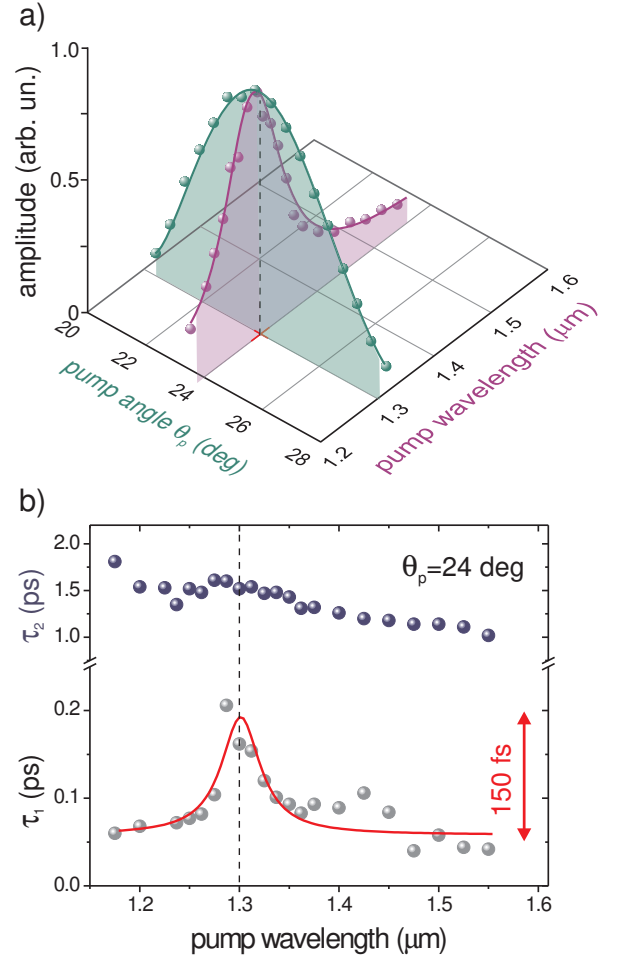}
    \caption{
    a) The amplitude of the pump-induced transmittance variations 
    is strongly increased at $\lambda_p=1.3$~$\mu$m, $\theta_p=24$~deg, corresponding to the SP excitation. The solid lines and shades are guides to the eye.
    b) Spectral dependence of the hot electrons transport ($\tau_1$, bottom) and thermalization ($\tau_2$, top) timescales.
    The solid red line indicates the pronounced enhancement of $\tau_1$ at the SP resonance ($\lambda_{\rm SP}=1.3$~$\mu$m).
    }
    \label{figs:param}
\end{figure}

Taking into account the above discussion on the relevant physical processes at this timescale, we fitted the double exponential function to the set of transmittivity data obtained at various pump wavelengths:

\begin{equation}\label{fitfunc}
   \Delta T(t)\propto A_1e^{-t/\tau_1}+A_2e^{-t/\tau_2}+A_3(1-e^{-t/\tau_2}).  
\end{equation}
%
%
This shape was convoluted with the Heaviside function $\Theta(t)$ with the finite step width set to account for the duration of the cross-correlation pump-probe signal $w\approx60$~fs measured independently. 
The first term $\propto A_1$ is related to the electron transport and describes spatial redistribution of the non-thermal electron population across the Au layer.
Indeed, immediately after the laser excitation, hot carriers (electrons and holes) are created, resulting in the alteration of the optical properties of the metal by virtue of the state-filling effect \cite{RazdolskiJPCM17}. Due to the spatial inhomogeneity of laser excitation, hot electron transport results in the in-depth equilibration of their distribution $n_N(z)$. Owing to the unequal in-depth sensitivity $s(z)$ of the probe beam to the state-filling effect, the hot electrons in-depth transport  results in transient variations of transmittance $\Delta T_N\propto\int_0^d n_N(z) s(z) dz$.
This spatial redistribution proceeds with its own characteristic time $\tau_1$ until the hot electron population is homogeneously distributed in Au. As such, $\tau_1$ can be estimated as a mean electron travel time $\tau_1\sim d/\langle v\rangle$ which in our case of ballistic transport yields $d/v_F\approx 50$~fs.

The second term in Eq.~\eqref{fitfunc} ($\propto A_2$) describes local relaxation of non-thermal electron distribution $n_N$ due to the electron
scattering. Here we
consider the interaction of the ensemble of hot electrons with a combined reservoir consisting of electrons and phonons in the thermal equilibrium (see Fig.~\ref{figs:over},right). As a result of this process, both the electron and lattice temperatures increase, which is accounted for in the third term of Eq.~\eqref{fitfunc} ($\propto A_3$). 
It results in an offset at large time delays since the neglected heat dissipation away from the laser spot occurs at much longer timescales.
%
This analytic function with two characteristic timescales $\tau_1, \tau_2$ provides excellent fits to the experimental transmittance traces, as seen in Fig.~\ref{figs:exp},b.

The data shown in Fig.~\ref{figs:exp},b exemplify the difference between the electron response at and away from the SP resonance, in particular, the enhanced variations $\Delta T$ at the SP resonance.
The blue data points
were obtained 
at $\lambda_p=1.5$~$\mu$m, where no phase-matched free-space SP excitation can be expected. From the fit procedure, we extract the times $\tau_1\approx 58$~fs, $\tau_2\approx 1.14$~ps, in a very good agreement with our above estimations.
%
In contrast, at the SP resonance $\lambda_p=1.3$~$\mu$m (Fig.~\ref{figs:exp},b, gray), we found $\tau_1\approx 162$~fs, $\tau_2\approx 1.52$~ps.
%
Systematic measurements with $\lambda_p$ varied
in the spectral vicinity of $\lambda_{\rm SP}$ allowed us to plot the resulting spectra of the amplitude of the variations $\Delta T/T(t)$ and $\tau_i$ in Fig.~\ref{figs:param}.
The amplitude peaks around $\lambda_{\rm SP}$, indicating the enhanced light-matter interaction, photon absorption and hot electron generation at the SP resonance (Fig.~\ref{figs:param},a). More interestingly, we found a consistent enhancement of the shorter timescale $\tau_1$ at the resonance, rising from $\approx 50$~fs up to $\approx 200$~fs (Fig.~\ref{figs:param},b). A similar increase expected in the spectral dependence of $\tau_2$ is concealed by the general non-resonant trend.
We attribute $\tau_1$ and $\tau_2$ to the characteristic timescales of the electron transport and thermalization, respectively.

We note that the excitation of SP cannot significantly modify the initial in-depth distribution of the hot electrons. Indeed, the SP electric field inside the metallic medium exponentially decays away from the interface with the characteristic length $\xi\approx22.7$~nm \cite{raether}
%
%
at $\lambda=1.3$~$\mu$m ($\varepsilon_m\equiv\varepsilon_m^{\prime}+i\varepsilon_m^{\prime\prime}=-77.2+6.8i$~\cite{JohnsonChristy}, $\varepsilon_d\approx5$~\cite{landolt}). This value is very close to the off-resonant penetration depth of the optical field $\zeta\approx23.5$~nm.
As such, the laser excitation profile in Au in the resonant ($\lambda_{\rm p}=1.3$~$\mu$m) and non-resonant cases are very similar, and the enhanced electron transport timescale $\tau_1$ cannot be explained by unequal initial population profiles of the primarily excited non-thermal hot electrons $n_N(z)$.

In order to explain this enhancement, we now briefly recall the conventional approach to the electron dynamics in metals.
The two temperature model \cite{Anisimov} describes the interaction of the electron and phonon reservoirs, 
where both are considered in the thermal equilibrium. In PMs, this picture is irrelevant on the sub-ps timescale, where non-thermal electrons are vital.
The electron dynamics is usually obtained from the rate equations, enabling the uncomplicated augmentation of the model with additional reservoirs such as non-thermal particles \cite{SunPRB94,RotenbergPRB09}.
In the case of sufficiently inhomogeneous laser excitation, the electron transport can be taken into account in the form of superdiffusion without the loss of generality.
Importantly, on the ultrashort timescale all other reservoirs can be omitted, as discussed above, and the dynamics of the hot non-thermal electrons $n_N(z,t)$ can be modeled with the following rate equation:

\begin{equation}\label{neqrate}
    \frac{\partial n_N}{\partial t}=-\frac{n_N}{\tau_2}+D^{\star}\frac{\partial^2 n_N}{\partial z^2}+S(z,t),
\end{equation}
where $D^{\star}$ is the generalized diffusion coefficient (in the purely diffusive limit $D^{\star}=\langle v\rangle l_e/3$), and $S(z,t)$ is the source term which describes the excitation of the hot electrons. This equation governs the dynamics of the non-thermal electrons on their own timescale $\tau_2$. The SP lifetime on a {\it continuous} Au/garnet interface at $1.3~\mu$m can be calculated \cite{raether,DerrienJOP16} as $\tau_{\rm SP}\approx310~{\rm fs}$.
%
%
%
%
Because $\tau_{\rm SP}\ll\tau_2$ (Fig.~\ref{figs:param},b), we can incorporate the SP excitation into Eq.~\eqref{neqrate} by modifying the source term $S$ into a convolution of the laser excitation and the SP dynamics:

\begin{equation}\label{spsource}
    S_{\rm SP}(z,t)=S(z,t)\ast Le^{-t/\tau_{\rm SP}}.
\end{equation}
Depending on the SP quality factor, the amplitude $L>1$ illustrates enhanced light absorption and hot electron generation in the metal due to the SP excitation. Here we neglect the spatial dependence of the SP-driven contribution $Le^{-t/\tau_{\rm SP}}$ due to the aforementioned similarity of the light penetration profiles in the resonant and non-resonant cases. Thus the role of collective electronic excitations is revealed: SPs act as delayed sources of hot non-thermal electrons, extending the time period when the hot electron dynamics (both local and non-local) is relevant. 
The impact of the SP excitation with the lifetime $\tau_{\rm SP}\sim0.1$~ps is more noticeable in the non-local dynamics occurring on a similar timescale $\tau_1$ than in the slower local electron thermalization ($\tau_2\sim1$~ps).

\begin{figure}[t]
    \centering
    \includegraphics[width=0.95\columnwidth]{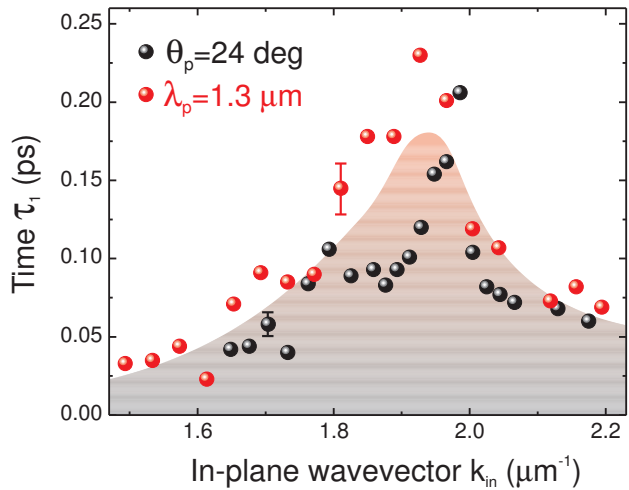}
    \caption{Consistency check for the spectral and angular probe of the time-resolved SP dynamics. The non-local relaxation time $\tau_1$ as a function of the in-plane pump wavevector component $k_{\rm in}$, obtained from the experiments with varied $\lambda_p$ (black, $\theta_p=24$~degrees) and $\theta_p$ (red, $\lambda_p=1.3$~$\mu$m). The remarkable agreement of the two datasets confirms the SP resonance-driven $\tau_1$ behaviour illustrated with the shaded area.
    }
    \label{figs:wavevector}
\end{figure}

The periodicity of the Au/garnet grating enables radiative SP losses which add up to the intrinsic (Joule) ones, reducing the real SP lifetime $\tau^{\prime}_{\rm SP}$. The data in Fig.~\ref{figs:param},b yields $\tau^{\prime}_{\rm SP}\approx 150$~fs, in line with a crude estimate $\tau^{\prime}_{\rm SP}\sim\lambda^2/c\Delta\lambda\approx120$~fs, where $\Delta\lambda\approx50$~nm is the spectral width of the SP resonances.
%
%
A pronounced SP resonance indicates the approximate balance of the Joule and unknown radiative losses \cite{raether}, decreasing the SP lifetime on a grating by a factor of two, as compared to $\tau_{\rm SP}\approx 310$~fs on a continuous interface, 
$\tau^{\prime}_{\rm SP}\approx\tau_{\rm SP}/2$.  An outstanding consistency in the determination of the SP lifetime
strongly corroborates our understanding of the SP role in non-thermal electron dynamics.

In order to unambiguously associate the observed effects with the SP excitation, in the second run of experiments we fixed the pump wavelength at $\lambda_p=1.3$~$\mu$m and varied the angle of incidence by rotating the sample, such that the angle between the probe and pump beams remained constant. Similarly to the previously described procedure, we fitted the function from Eq.~\eqref{fitfunc} to the measured time-resolved transmittivity traces and extracted the fit parameters. Then, taking advantage of the interchangeability of $\lambda_p$ and $\theta_p$
for the phase-matched SP excitation,
we plotted the hot electron transport timescale $\tau_1$ as a function of the in-plane momentum $k_{\rm in}=k_0\sin\theta_p$. We found a great degree of similarity between these sets of data obtained in the two independent measurements (Fig.~\ref{figs:wavevector}).  Remarkably, the only crossover of the two clearly distinct experimental approaches is the phase-matched SP excitation at $k_{\rm in}=k_{\rm SP}-mk_G$, thus unequivocally corroborating our understanding of the collective dynamics of non-thermal electrons. Within our framework, the data obtained in the two distinct series of measurements
demonstrate increased hot electron lifetime at the SP resonance and elucidate the role of SPs in hot electron dynamics.

Summarizing, we have performed time-resolved measurements of the hot electron dynamics in Au on the sub-picosecond timescale in the vicinity of the SP resonance. Systematic measurements with two independently varied parameters allowed us to unambiguously associate the apparent retardation of the hot electron population decay with the SP excitation. An estimation of the SP lifetime at the Au/dielectric garnet interface of about $150$~fs can be inferred from the pronounced enhancement of the relevant hot electrons lifetime at the resonance.
In practice, the lifetime of non-thermal electron population can be key for the duration of the spin-polarized current pulses \cite{AlekhinPRL17,RazdolskiNatComm17,LalieuPRB17}, spin Seebeck \cite{KimlingPRL17} and inverse spin-Hall \cite{SeifertNatComm18} effects in metal-dielectric bilayers, or laser-induced demagnetization rate \cite{EschenlohrNatMater13,BergeardPRL16}. Allowing for external engineering of their losses, SPs can be envisioned as flexible photonic tools for the spintronic functionality, as well as other applications of the hot electron science and technology \cite{MukherjeeNanoLett13,BrongersmaNatNano15}. Furthermore, we have shown how the time-resolved transmittivity data can be employed for the direct measurement of SP lifetimes at arbitrary corrugated interfaces with unknown losses. 
Our results significantly advance the understanding of collective hot electron dynamics and clearly identify the role of SP resonances in the ultrafast optical response. We envision further crossover of ultrafast photonics with spin- and electronics for the efficient control of electron dynamics in metal-dielectric nanostructures.

\acknowledgments

The authors thank A.~Melnikov and A.~Kirilyuk for valuable comments, and M.~Wolf, T.~V.~Murzina, and A.~Maziewski for their support. This work has been funded by the National Science Centre
Poland (Grant DEC-2017/25/B/ST3/01305).


%

\end{document}